\documentclass[sigconf]{acmart}
\newcommand{\myname} {{\em RLIRank}}
\AtBeginDocument{%
  \providecommand\BibTeX{{%
    \normalfont B\kern-0.5em{\scshape i\kern-0.25em b}\kern-0.8em\TeX}}}

\setcopyright{iw3c2w3}
\copyrightyear{2020}
\acmYear{2020}
\acmDOI{10.1145/3366423.3380047}

\acmConference[WWW '20]{Proceedings of The Web Conference 2020}{April 20--24, 2020}{Taipei, Taiwan}
\acmBooktitle{Proceedings of The Web Conference 2020 (WWW '20), April 20--24, 2020, Taipei, Taiwan} 
\acmPrice{}
\acmISBN{978-1-4503-7023-3/20/04}

\begin{document}

\title{RLIRank: Learning to Rank with Reinforcement Learning for Dynamic Search}

%

\author{Jianghong Zhou}
\affiliation{%
  \institution{Emory University}
  \city{Atlanta}
  \state{USA}
}
\email{jianghong.zhou@emory.edu}

\author{Eugene Agichtein}
\affiliation{%
  \institution{Emory University}
  \city{Atlanta}
  \country{USA}}
\email{eugene.agichtein@emory.edu}

%

%
\begin{abstract}
To support complex search tasks, where the initial information requirements are complex or may change during the search, a search engine must adapt the information delivery as the user's information requirements evolve. To support this dynamic ranking paradigm effectively, search result ranking must incorporate both the user feedback received, and the information displayed so far. To address this problem, we introduce a novel reinforcement learning-based approach, \myname. We first build an adapted reinforcement learning framework to integrate the key components of the dynamic search. Then, we implement a new Learning to Rank (LTR) model for each iteration of the dynamic search, using a recurrent Long Short Term Memory neural network (LSTM), which estimates the gain for each next result, learning from each previously ranked document. To incorporate the user's feedback, we develop a word-embedding variation of the classic Rocchio Algorithm, to help guide the ranking towards the high-value documents. Those innovations enable \myname{ } to outperform the previously reported methods from the TREC Dynamic Domain Tracks 2017 and exceed all the methods in 2016 TREC Dynamic Domain after multiple search iterations, advancing the state of the art for dynamic search.  
\end{abstract}

%
%

%

%

%
\maketitle


\section{Introduction}
 In complex search tasks, such as prior art patent search, users require both high relevance of the documents returned, and the coverage of multiple perspectives on the topics. Therefore,   users may continue interacting with the search engine. Simultaneously, the search engine should be able to mine the latent intents of the users from the interactions \cite{yang2016trec}. This is the basic scenario of dynamic search. 

\begin{figure}[h]
  \centering
  \includegraphics[width=\linewidth]{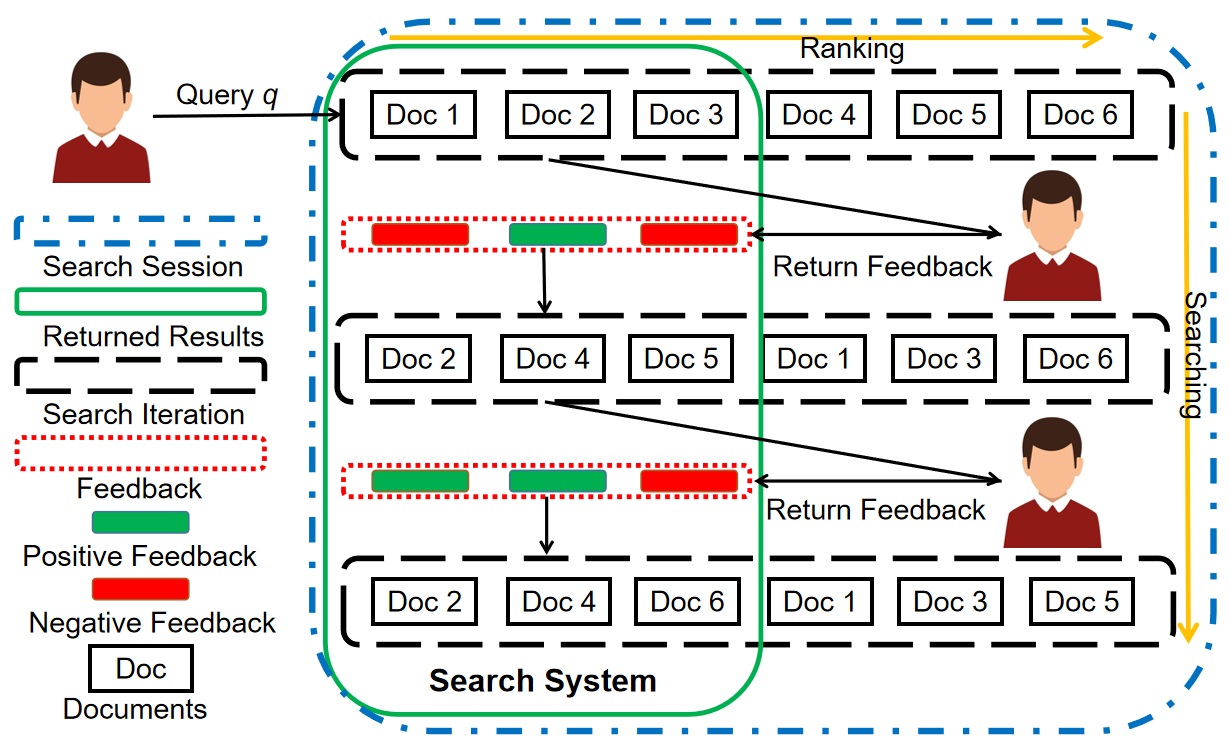}
  \caption{Flow of Dynamic Search with user feedback. The returned results are dynamically re-ranked based on the user's feedback. }
  \Description{The TREC domain Search Process.}
\end{figure}

Figure 1 illustrates the dynamic search process. A dynamic search includes a search session, containing many search iterations, whose results are a list of documents. Each search iteration includes many search units. The returned result of each search unit is a document. During the whole search session, users provide one query to the search engine, and provide the feedback after each iteration \cite{kanoulas2018overview}. 

We formulate our approach based on the TREC Dynamic Domain setting \cite{yang2016trec}. In this setting, when a search session begins, a user searches for a query with several subtopics, which are not exposed to the search engine. This is the first search iteration. After that, the search engine analyzes the query and the documents' passages to rank the documents. Then, the search results are returned to the user, which is simulated in the TREC dataset. 
This simulator sends feedback to the search engine, indicating the relevance scores of each subtopic with the query. However, the feedback does not contain any information about the content or number of the subtopics. Based on the feedback, the search engine computes a score to evaluate this search iteration's quality, such as $\alpha$-DCG \cite{clarke2008novelty}. After receiving the feedback, if search session is not suspended, the search engine returns the new results based on the feedback. When reaching the stop conditions, the search session stops.

Based on this setting, extensive prior research studied the dynamic ranking mechanisms. In 2016 TREC Dynamic Domain track, \cite{albahem2016rmit} introduced four models. The highest performing model was {\bf rmit-lm}, built by the language modeling approach from Apache Solr \cite{grainger2014solr}. Another noteworthy model is {\bf rmit-lm-oracle-1000}, which operates by retrieving the first 1000 documents, removing the irrelevant documents from the list using the ground truth, and returning the top 5 documents. This model was the state of art in 2016 TREC Dynamic Domain track. However, this model uses ground truth labels to filter out documents, which is usually not available in realistic tasks. 
Based on 2016 TREC Dynamic Domain, \cite{yang2017contextual} applies the contextual bandit approach to dynamic search. However, the method does not outperform the state of art in 2016 TREC Dynamic Domain and lacks basic documents ranking metrics in the experiments.  In 2017 TREC Dynamic Domain track, more models were proposed to improve the overall quality of the search session. 
Among those methods, {\bf ictnet-params2-ns} is the state of art. Although those models present encouraging progress in dynamic search, they are suffering some common flaws. Most of the aforementioned methods largely depend on some language models, and traditional supervised Learning to Rank (LTR) models.
The reported models were trained by pointwise, pairwise or listwise methods. However, these approaches fail to capture all the ranked documents' information to improve the overall quality of the search session. 

To address those problems, we propose a new method -- \myname{}: Learning to Rank with Reinforcement Learning for Dynamic Search. \myname{ } applies a reinforcement learning paradigm to conduct each search session of the dynamic search \cite{sutton2018reinforcement}.  For each search iteration, we apply a stepwise training method to tune a stacked recurrent neural network, which explores and returns the search results \cite{sutskever2014sequence}.  We also apply an embedding-oriented Rocchio algorithm to digest the feedback \cite{joachims1996probabilistic}. The experiments on TREC 2016 Dynamic Domain and TREC 2017 Dynamic Domain show the \myname{ } is an effective dynamic search model, which outperforms the previous methods by up to 6.2\%.  

In summary, our contributions include:
\begin{itemize}
    \item An effective implementation of the deep reinforcement learning paradigm adapted to the dynamic search problem, which captures the relationship between the user feedback and the document content.
    \item An effective deep value network model architecture, which incorporates a stacked LSTM and the associated new stepwise method to train the RLIRank model. 
    \item Adaption of the classical Rocchio feedback algorithm to the deep RL framework which extends the classic Rocchio algorithm to use dense word embeddings for relevance feedback in the neural ranking setting. 
    \item Evaluation of the RLIRank method on the 2016\&2017 TREC Dynamic Domain datasets, showing that \myname{ } outperforms all previously published methods in these two tracks, after as few as 5 search iterations.
\end{itemize}


\section{Related Work}
We briefly review prior work related to \myname{ }, including basic Learning to Rank methods, as well as the more recent research on applying reinforcement learning and deep learning for the documents ranking problems or dynamic search problems.

\subsection{Learning-to-Rank (LTR) methods}


In previous research summarized in \cite{liu2009learning}, most of methods to solve Learning to Rank problems are pointwise, pairwise or listwise approaches. The pointwise approaches use a single document as its input in learning and define the loss functions by the relevance of each document. In contrast, pairwise approaches take document pairs as instances, while listwise approaches consider the whole list of documents \cite{lin2007ordinal}\cite{li2008mcrank}\cite{lee2014large}\cite{burges2005learning}\cite{xu2007adarank}\cite{cao2007learning}. 


Although those approaches greatly model different basic features of LTR, those methods are still restricted to a certain data structure. In fact, during the search process, the length of the rank is increasing. None of the aforementioned method models this important dynamic process. Inspired by this, we further extend the listwise method to the stepwise method.

\subsection{Reinforcement Learning-based Methods}



Recently, many approaches to LTR have been proposed based on reinforcement learning. The early research on applying reinforcement learning in the ranking is  \cite{monahan1982state}, which uses a Partially Observable Markov Decision Processes (POMDPs).  In 2013, Guan et al. further developed it to a session search algorithm, which improves the performance of the MDP by attaching different importance to different time documents in the reward function \cite{guan2013utilizing}. Expanding on these two methods, Glowaka et al. designed a reinforcement system supporting active engagement of users during search \cite{glowacka2013directing}. 

Most recently, and closest to our work, is a recently introduced method which uses Reinforcement Learning to Rank with Markov Decision Process (MDP)\cite{wei2017reinforcement}. One of the central ideas of this method is treating the document list as a sample, instead of evaluating documents one by one. This allows MDP to optimize performance evaluation metrics, like NDCG score, directly. 
As a result, the MDP method was demonstrated to be the new state of the art on the MQ2007 and MQ2008 benchmarks. In follow-up work, \cite{feng2018greedy}\cite{liu2018novel} successfully applied MDP to search diversification problems.

\subsection{Neural Information Retrieval: Deep Learning for Ranking}
Deep learning is one of the most active research areas in recent years \cite{schmidhuber2015deep}. The great success of deep learning in image recognition and natural language processing (NLP) attracts many researchers to explore possible utilization of deep learning methods to other applications \cite{lecun2015deep}\cite{collobert2008unified}.

There have been many research efforts on using deep learning for Learning to Rank problem. For example, \cite{severyn2015learning} uses a convolution neural network (CNN) to rank short text pairs. \cite{wang2014learning} proposes a CNN-based method to rank image files. Additionally, \cite{santos2015classifying} applies a neural network to discover the relation for ranking.  Those methods solve the problem effectively, but they do not naturally fit the structure of the dynamic search.



. 



\begin{figure}[htbp]
\centering
    \begin{minipage}[t]{0.47\textwidth}
    \centering
    \label{fig:subfig:b} 
    \includegraphics[width=\linewidth]{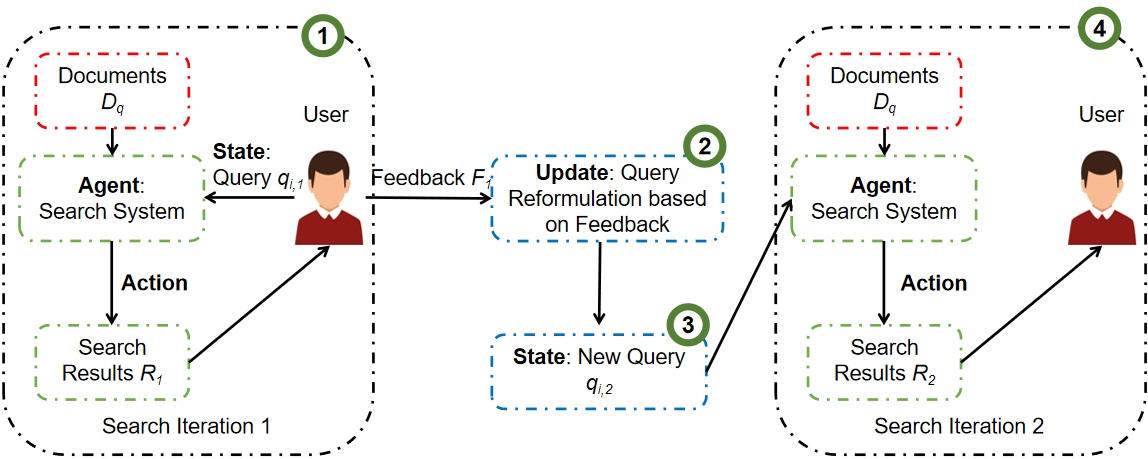}
\end{minipage}
\caption{The framework of the dynamic search model and \myname. }
  \Description{\myname{ } reinforcement framework. }
\end{figure}

\section{The \myname{ }Framework}

In this section, we apply reinforcement learning to model the important steps of dynamic search. As Figure 2 illustrates,  the search system is  \textit{Agent}. The ranking process is an \textit{Action}. The feedback from the users is the \textit{Reward}. The details of the framework are summarized as follows:

 \noindent\textbf{\textit{Agent}}:
 The \textit{Agent} is the search system. The \textit{Agent} ranks the documents based on the following policy:
  \begin{equation}
    \pi(a|S,\phi) = \left\{
\begin{array}{rcl}
\pi_1 & & {{\rm if}\ p<\epsilon}\\
\pi_2 & & {{\rm if}\ p\geq \epsilon}
\end{array} \right.
 \end{equation}
 Where $p$ decides the policy. $\epsilon$ is the threshold of the greedy strategy.
 When the \textit{Agent} chooses $\pi_1$, the \textit{Action} returns a random document. If the \textit{Agent} chooses $\pi_2$, then the \textit{Agent} returns a document based on the probability calculated by Equation (2).
   \begin{equation}
    Pr(a|S,\phi)=\frac{V_\phi (f(S,a))}{\sum_{a_i \in A}V_\phi(f(S,a_i) )}
 \end{equation}
 Where  $\phi$ represents all the parameters of the deep value network $V_{\phi}$, which is introduced in Section 4.2. $S$ is the current \textit{State}, $a$ is the current action. $f(S,a)$ is the transition function that updates \textit{State} $S$ based on the action $a$. 
 
  \noindent\textbf{\textit{State}}:
  \textit{State} $S$ represents the current system's search context, which includes two parts:
   \begin{equation}
     S = [R,q]
 \end{equation}
 Where $R$ is a list of ranking documents and $q$ is the query. Since \textit{State} contains both $R$ and $q$, two kinds of actions are required to update \textit{State}.
 
   \noindent\textbf{\textit{Actions}}:
   The first action, which we note it as $a_r$, is the intermediate action. In Figure 1, to generate a list of ranking documents in a search iteration, we first rank each document. $a_r$ represents this ranking process. Based on the Policy (1), with \textit{Action} $a_r$, if we choose a document, we denote it as $d_r$.
   
 The other action updates the query after each search iteration, which is noted as $a_t$. With the \textit{Action} $a_r$, we denote the new query as:
  \begin{equation}
     q_t = T(q,F)
 \end{equation}
 Where $T$ is the feedback function to reformulate a new query based on the original query $q$ and the user feedback $F$, which is further illustrated in Section 4.3.
 
  \noindent\textbf{\textit{Transition Function}}: \textit{Transition Function} $f$ defines how the search system is transited to a new \textit{State}. For different actions, the transition function works in different ways. Basically, the transition function pairs the documents and the queries and update \textit{State}. We define the transition function $f$ as:
   \begin{equation}
     f(S,a_r) = \{(d_1,q),(d_2,q),(d_3,q),...,(d_k,q),(d_r,q)\}
 \end{equation}
    \begin{equation}
     f(S,a_t) = \{(d_1,q_t),(d_2,q_t),(d_3,q_t),...,(d_k,q_t)\}
 \end{equation}
In Equation (5), \textit{State} is updated by appending the new document to the ranked list. In Equation (6), State is updated by replacing the query with the new formulated query. The result of $f(S,a)$ is an ordered set of the document and query pairs, which is the input for the ranker.

\noindent\textbf{\textit{Rewards}}: 
 In different steps of the dynamic search, the \textit{Rewards} from the search system are varied. Although the most obvious \textit{Reward} is the feedback from the user, the relevance scores available in the training process should not be ignored. During the training process, the relevance scores between the query and the document are available. The feedback, which is the relevance scores among different subtopics of the query and the documents are only accessible after a list of ranked documents is given. However, such a \textit{Reward} is evaluated by pairing the new query and the document. Therefore, after ranking a document, we evaluate the action based on the given relevance score in the training process. We evaluate the action with different metrics, such as NDCG, $\alpha$-NDCG and et al. We denote the \textit{Reward} $R$ as: 
 \begin{equation}
 R=V_* (f(S,a))  
  \end{equation}
 To utilize intermediate \textit{Rewards} to rank the next document better after, we minimize the following loss function after each action:
    \begin{equation}
     L(S,a|\phi)=\min_{\phi}\left(V_{\phi}(f(S,a))-V_{*}(f(S,a))\right)^2
 \end{equation}

 Where $\phi$ is the parameters of the deep value network $V_{\phi}$. This loss function evaluates the distance between the real metric and the estimated metric. The real metric requires the relevance scores between the query and the document, yet the estimated metric or deep value network does not require the scores, which is important for the testing process or the unsupervised situation.


\section{The \myname{ }Implementation}

The \myname{ } implementation includes three parts. 
First, we introduce a stepwise learning framework, which is designed to train the value network $V_\phi$. In Section 4.2, we describe the configuration of \myname{ } deep value network $V_\phi$. Finally, we introduce our definition of feedback function $T$. This approach incorporate user feedback and the last search iteration query to formulate a new one, which derives from the classic Rocchio algorithm.

\subsection{The Stepwise Learning Framework}
In this section, we create a new approach --  the stepwise learning framework to train the deep value network $V_\phi$.

The details of this learning framework are illustrated in Figure 3 (a). First, a document is chosen from the candidate list. When the document is chosen, it is also removed from the candidate list. The document choosing process depends on a score from $V_\phi$ and the $\epsilon$-greedy strategy. The $V_\phi$ is a model with multiple entries, such as RNN, LSTM, and et al. The number of their activated entries is determined by the number of the chosen documents. 
\begin{figure*}[htb]
\centering
\[
\begin{array}{cc}
    \includegraphics[width=220pt]{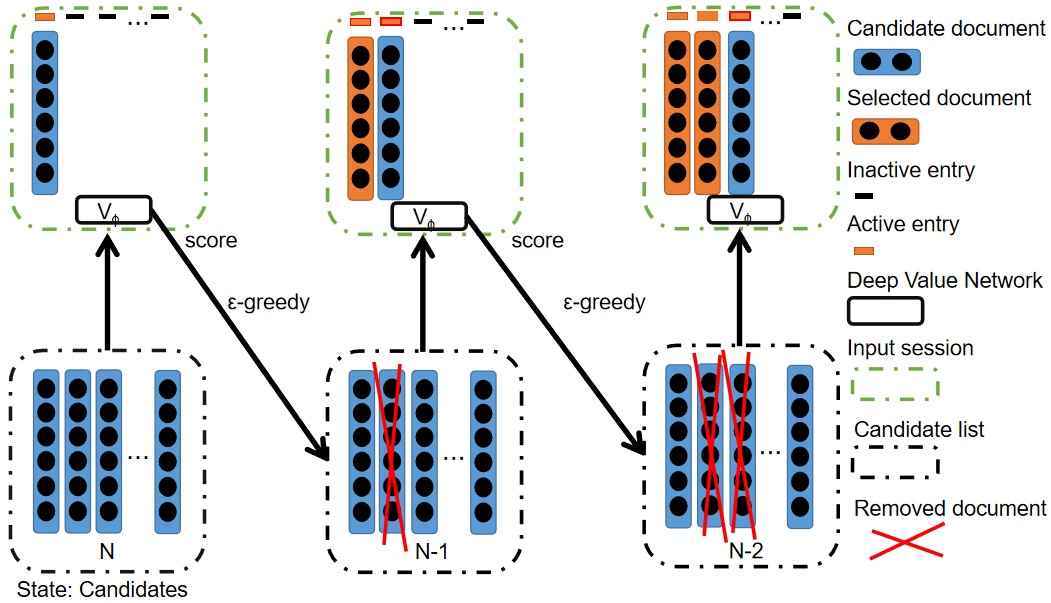}
  &
 \includegraphics[width=220pt]{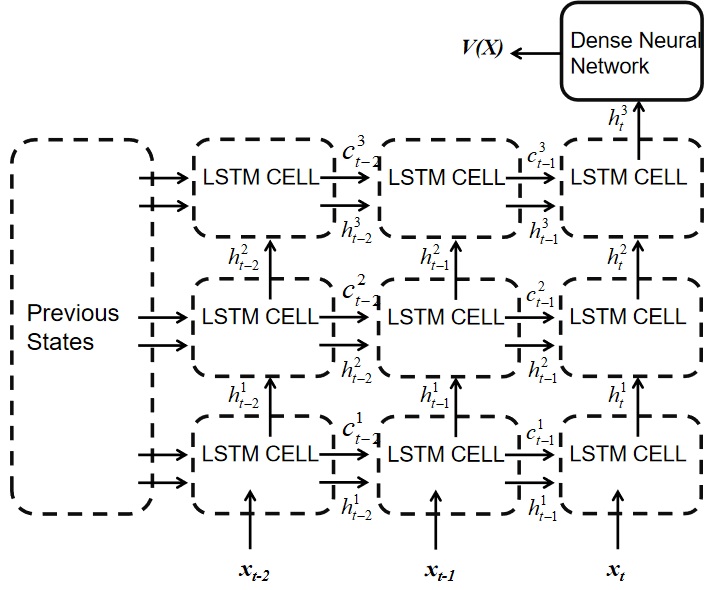}
  \\
  (a) & (b) 
\end{array}
\]
\vspace{-3mm}
\caption{Illustration of the stepwise learning framework (a); The structure of the stacked LSTM. (b)}
\label{fig:rli-rank-results}
\end{figure*}




\subsection{Implementation of $V_\phi$ Network}

In this paper, we configure the $V_\phi$ with a stacked LSTM, which is introduced with three key parts: Input, Stacked LSTM Model Architecture and Loss Function. 

\noindent\textbf{Input:} The input of the proposed stacked LSTM is dynamic because we apply a stepwise learning framework to train the stacked LSTM.  We further justify the input as follows: Selecting the position: $X=X'\cup \{x_t\}$. $X'$ is an order set of ranked items, $x_t$ is the ranking item, $\cup$ means appending $x_t$ to $X'$. From Figure 3(b), we find that the $x_t$ is considered as the most recent input item of the LSTM. In the list of the ranked document, the nearer to the position $t$ the item is, the more recent it is. The input of each cell $x$ is an integration of a document and a query. For instance, if we use a vector $d$ to represent  the document, and use a vector $q$ to represent the query, then $x = [d,q]$. Therefore, $x$ is a query and document pair.

\noindent\textbf{Stacked LSTM Model Architecture:}
The construction of this stacked LSTM is based on the demand of the experiment. The selection of the depth of the model is a trade-off between computation and performance. Based on the experiment and analysis in Section 6, we build a three-layer stacked LSTM as shown in Figure 3(b).
 The stacked LSTM contains some gate functions, allowing the cells to transfer the long term information or forget unnecessary information. The forget gate of $j$th layer, $k$th input $j_k^j$, the input gate $i$, the output gate $o$, the cell state $c$ and the hidden state $h$ of $V_\phi$ is computed as follows:
\begin{displaymath}
     f_k^j=\sigma(W_f^{j}h^{j-1}_{k}+U_f^{j}h_{k-1}^j+b_f^j),
\end{displaymath}
\begin{displaymath}
    i_k^j=\sigma(W_i^{j}h^{j-1}_{k}+U_i^{j}h_{k-1}^j+b_i^j),
\end{displaymath}
\begin{displaymath}
    o_k^j=\sigma(W_o^{j}h^{j-1}_{k}+U_o^{j}h_{k-1}^j+b_o^j),
\end{displaymath}
\begin{displaymath}
    c_k^j=f_k^j\circ c^j_{k-1}+i^j_k\circ \tanh(W_c^j h^{j-1}_{k}+U_c^{j}h_{k-1}^j+b_c^j),
\end{displaymath}
\begin{displaymath}
    h_k^j=o_k^j\circ \tanh(c_k^j),
\end{displaymath}
in which, $\sigma(x)=\frac{1}{1+e^{-x}}$ and applies to every entry. $h^0_k=x_k$, the operator $\circ$ denotes the element wise product. The values of $c_0^j$ and $h_0^j$ are assigned by the Glorot Uniform Initializer in our paper. The weight metrics and the bias vectors: $W_f^{j}$,$W_i^{j}$,$W_o^{j}$,$W_c^{j}$,$b_f^j$,$b_i^j$,$b_o^j$, and $b_c^j$ are also assigned by the same initializer.

The output of the LSTM part is $LSTM(X',x_t)=h_K^J$, $K$ is the ranking position, and $J$ is the top layer's label.


\noindent\textbf{Loss Function:} 
The loss function is Equation (8).

\subsection{An embedding Rocchio algorithm}

The embedding Rocchio algorithm is the feedback organizer or the function $T$ in Equation (4). 


In \myname{ }, to digest the feedback from the users and make it observable for the \textit{Agent}, we reformulate the query after each search iteration. The new query further optimizes the $V_\phi$.  

However, the query reformulation leads to an important issue. If we directly add words from relevant documents or remove them from the query, like the classic Rocchio algorithm, the variation is discrete, which easily results in unexpected results. Besides, the action of removing the words may be useless,  because the query is so short that possibly does not contain the words. 

To resolve the aforementioned problem, we further improve the classic Rocchio algorithm to an embedding Rocchio algorithm. This algorithm generates a new query based on the list of the ranked documents and the feedback. It is summarized as follows:
\begin{equation}
    q_{n+1}=(1-\gamma^n(b-c))q_n+\gamma^n (\frac{b \sum_{d_j\in D^n_r}{d_j}}{|D^n_r| }-\frac{c \sum_{d_k\in D^n_{nr}}{d_k}}{|D^n_{nr}| })
\end{equation}
Where $q_n$ is the encoded query for the search iteration $n$. $D^n_r$ is the documents receiving positive feedback from the users in the search iteration $n$. $D^n_{nr}$ is the documents not receiving feedback or receiving negative feedback from the users in the search iteration $n$. $d$ is the encoded documents. $\gamma$ is the discount rate of different search iteration. $b$ and $c$ are the weights of the positive feedback and negative feedback respectively.

\section{Experimental Setting}
In this section, we first introduce the basic experimental settings and datasets, then we experimentally compare \myname{ } with different methods to report the main results. 


\subsection{Benchmark Datasets}
There are two main datasets we consider: 2016 TREC Dynamic Domain (DD), 2017 TREC DD. 
They share similar datasets' settings. 2016 TREC DD contains Polar domain and Ebola domain. The Polar domain contains 26 queries and 1741530 documents. The Ebola domain contains 27 queries and 682157 documents. 2017 TREC DD dataset is the archives of the New York Times in 20 years, which has 60 queries and 1855658 documents \cite{sandhaus2008new}. Each query has several subtopics, and the documents have a relevant score for some subtopics provided by a {\em Jig} user simulator. The relevant score of a document to a query is the summation of all the scores of the subtopics. 
Although the datasets have a similar setting, the 2016 DD track is $\alpha$-DCG as the primary metric, and the 2017 uses $nSDCG$ \cite{hagen2010webis}\cite{clarke2008novelty}.

\textbf{RLIRank settings}: The implementation of the proposed models is based on Tensorflow's basic models, and the parameters not mentioned are mostly the default settings. The stacked LSTM is constructed by 3 layers LSTM, 5 iterations, and a dense neural network. The dense neural network consists of 5 hidden layers and 1 softmax function. The neurons' numbers for LSTM cells or hidden layers are varied based on the experiments.  The hidden neurons' number for each hidden layer is 1024, 512, 256, 16, 8, and the input size of the LSTM cell is 1024. The $V_*$ of LSTM  is $\alpha$-DCG for 2016 TREC DD, and DCG for 2017 TREC DD. We apply the default Google Universal Sentence Encoder to transfer the words to the vectors. We encode both the query and document passage's content into a 512 vector and then integrate them into an input unit -- a 1024 vector. The reason to use the universal encoder is that the value space of the encoded vectors is dense, which is suitable to apply the embedding Rocchio algorithm. Through little sample dataset test, we set $\gamma$ of the embedding Rocchio algorithm is 0.9, $b$ is 0.75, and $c$ is 0.25. During the training process, $\epsilon=0.5$. After each 1000 epochs, $\epsilon_{new}=\epsilon_{old}\times 0.9$. In testing process, $\epsilon = 0$. The drop-off rate of the neural network is 0.5, and all the activate functions in the neural network are ReLU. The model's training is stopped when the training improvement is less 0.01\%. 

\subsection{Baselines}
The baselines methods include the high-performing models appearing in the 2016 TREC Dynamic domain and 2017 TREC Dynamic Domain.

\begin{itemize}
    \item \textbf{rmit-oracle-lm-1000:} The model firstly retrieves 1000 documents using Solr with the content language model. Then they use the ground truth to remove irrelevant documents from the initial list of documents. For each iteration, it
returns the next 5 relevant documents from the initial list \cite{albahem2016rmit}.
    \item \textbf{rmit-lm-nqe:} This method uses the Language modeling approach as
implemented in Apache Solr using Dirichlet smoothing and default parameters.
No query expansion (nqe) was applied \cite{albahem2016rmit}.
    \item \textbf{ufmgHS2:} Hierarchical diversification with single-source subtopics and cumulative stopping condition \cite{moraes2016ufmg}.
    \item \textbf{ufmgHM3:} Hierarchical diversification with multi-source subtopics and window based stopping condition \cite{moraes2016ufmg}.
    \item \textbf{ictnet-emulti:} The model uses xQuAD and query expansion \cite{zhang2017ictnet}.
    \item \textbf{ictnet-params2-ns:} The model is the same as \textbf{ictnet emulti}, but changes parameters of other solutions. Not use stop strategy \cite{zhang2017ictnet}. 
    \item \textbf{dqn-5-actions:} Use DQN to choose 5 possible search actions \cite{li2018intelligent}.
    \item \textbf{galago-baseline:} The first 50 results returned by galago \cite{cartright2012galago}.
    \item \noindent\textbf{\myname:} \myname{ } Our method described above.
\end{itemize}

The five-fold validation method is applied to all the experiments. All baseline models are the most recent models or high-performance models.  To confirm the significance of the improvements, we applied statistical significance testing (two-tailed Student's t-test) which reports p-values of < 0.05 for significant improvements.

\section{Results and discussion}

\begin{table*}
\small
  \caption{$\alpha$-NDCG of baselines, and \myname{ } on 2016 TREC Dynamic Domain dataset and nSDCG of ictnet-params2-ns, ictne-emulti, galago-baseline, dqn-5-actions, clip-baseline, and \myname{ } on 2017 TREC Dynamic Domain dataset. The best  performance results are highlighted in bold font. Results marked with * indicate significant improvements with p<0.05.}
  \label{tab:commands}
  \begin{tabular}{c|ccccc|cccl}
  \multicolumn{1}{c}{ }&\multicolumn{5}{c}{$\alpha$-NDCG of 2016 TREC Dynamic Domain}&\multicolumn{4}{c}{nSDCG of 2017 TREC Dynamic Domain}\\ 
    \toprule
    Iteration&\myname &rmit-oracle.lm.1000&ufmgHM3&ufmgHS2&rmit-lm-nqe&\myname&ictnet-params2-ns&galago-baseline&dqn-5-actions\\
    \midrule
      1&0.4947&\bf{0.6874}&0.3516&0.3516&0.3581 &\bf{*0.6517}&0.4545&0.4337&0.4457\\ 
     2&0.6055&\bf{0.7116}&0.4055&0.4079&0.3908 &\bf{*0.6517}&0.4902&0.4590&0.4761\\ 
     3&0.6500&\bf{0.7251}&0.4306&0.4256&0.4073 &\bf{*0.6497}&0.4971&0.4594&0.4805\\ 
      4&0.7224&\bf{0.7346}&0.4377&0.4335&0.4250 &\bf{*0.6483}&0.4943&0.4526&0.4819\\ 
      5&\bf{0.7396}&0.7387&0.4417&0.4360&0.4322 &\bf{*0.6477}&0.4982&0.4506&0.4837\\ 
      6&\bf{0.7445}&0.7406&0.4439&0.4366&0.4419 &\bf{*0.6488}&0.4954&0.4519&0.4841\\ 
      7&\bf{*0.7657}&0.7438&0.4446&0.4367&0.4461 &\bf{*0.6491}&0.4932&0.4548&0.4837\\ 
      8&\bf{*0.7685}&0.7448&0.4458&0.4367&0.4515 &\bf{*0.6492}&0.4943&0.4564&0.4857\\ 
      9&\bf{*0.7871}&0.7452&0.4468&0.4367&0.4555 &\bf{*0.6497}&0.5003&0.4592&0.4853\\ 
     10&\bf{*0.7927}&0.7468&0.4481&0.4368&0.4584 &\bf{*0.6499}&0.5033&0.4581&0.4850\\ 
    \bottomrule
  \end{tabular}
\end{table*}

\begin{table*}
\small
  \caption{$\alpha$-NDCG of baselines, and \myname{ } on 2016 TREC Dynamic Domain dataset and nSDCG of ictnet-params2-ns, ictne-emulti, galago-baseline, dqn-5-actions, clip-baseline, and \myname{ } on 2017 TREC Dynamic Domain dataset. The best  performance results are highlighted in bold font. Results marked with * indicate significant improvements with p<0.05.}
  \label{tab:commands}
  \begin{tabular}{c|ccc|ccc}
  \multicolumn{1}{c}{ }&\multicolumn{3}{c}{$\alpha$-NDCG of 2016 TREC Dynamic Domain}&\multicolumn{3}{c}{nSDCG of 2017 TREC Dynamic Domain}\\ 
    \toprule
    Iteration&\myname &rmit-oracle.lm.1000&ufmgHM3&\myname&ictnet-params2-ns&galago-baselin\\
    \midrule
      1&0.4947&\bf{0.6874}&0.3516 &\bf{*0.6517}&0.4545&0.4337\\ 
     
     3&0.6500&\bf{0.7251}&0.4306 &\bf{*0.6497}&0.4971&0.4594\\ 
     
      5&\bf{0.7396}&0.7387&0.4417 &\bf{*0.6477}&0.4982&0.4506\\ 
      
      7&\bf{*0.7657}&0.7438&0.4446 &\bf{*0.6491}&0.4932&0.4548\\ 
      
     10&\bf{*0.7927}&0.7468&0.4481 &\bf{*0.6499}&0.5033&0.4581\\ 
    \bottomrule
  \end{tabular}
\end{table*}
Table 1 shows that the improvement in the proposed \myname{ } is substantial.  Compared to the current state of art model, \myname{ } improves up to 6.1\% on the TREC 2016, and by over 20\% on the TREC 2017. 

We observe, on the TREC 2017 Dynamic domain, that \myname{ } has a high first iteration performance. Since the performance of the other search iterations depends on the first iteration, the first iteration results are significant. The first iteration search is a pure LTR problem. Therefore, the improvement of the first search iteration results from the proposed deep value network.


 We first explore the effect of the numbers of the stacked LSTM's layers. The results are presented in Figure 4. We find that as layers of the stacked LSTM increases, the performance improves. However, the improvement of the model decreases drastically when the number of layers is bigger than 3. To compromise the computation and the performance, we choose a 3-layer stacked LSTM as our deep value network.
We further compare our proposed deep value network $V_\phi$ with the traditional listwise methods and an MDP method on the standard LTR benchmarks. This MDP method was the state of art of this benchmark.
Table 2 shows that the improvements in \myname{ } over other methods are substantial. Compared to the current reinforcement state of the art, MDP, \myname{ } improves between 5.9\% and 8.4\% on the MQ2007 benchmark, and by 15\%-19\% on the MQ2008 benchmark. 
\begin{table*}
\small
  \caption{NDCG of RankSVM, ListNet, AdaRank-NDCG, MDP, and \myname{ }\  on MQ2007 and MQ2008 dataset. The best performance results are highlighted in bold font. The results marked with * indicate significant improvements with p<0.05.}
  \label{tab:commands}
  \begin{tabular}{c|ccccc|ccccl}
  \multicolumn{1}{c}{ }&\multicolumn{5}{c}{MQ2007}&\multicolumn{5}{c}{MQ2008}\\ 
    \toprule
    Method&NDCG@1 &NDCG@2&NDCG@3&NDCG@4&NDCG@5&NDCG@1 &NDCG@2&NDCG@3&NDCG@4&NDCG@5\\
    \midrule
    RankSVM&{.410}&.407&.406&.408&0.414&.363&.399&.429&.451&.470 \\
    ListNet &.400&.406&{ .409}&.414&.417&.375&.411&.432&.457&.475\\ 
AdaRank &.388&.397&.404&.407&.410&.383&.421&.442&.465&.482\\
MDP &0.404&{ .408}&.408&{ .416}&{ .419}&{ .409}&{ .435}&{ .463}&{ .481}&{ .510}\\ 
\myname &{\bf *.434}  &{\bf *.440}  &{\bf *.444}&{\bf *.446}&{\bf *.451}&{\bf *.487}&{\bf *.520}&{\bf *.547}&{\bf *.568}&{\bf *.587}\\
    \bottomrule
  \end{tabular}
\end{table*}

The results prove that ranking model of \myname solves the traditional LTR problems well and thus promises the high performance of \myname{ } in dynamic search problems. We attribute the improvements of it to the fact that while most of the baseline algorithms compared are list-based, the stacked LSTM is trained by more completed information because of the stepwise training. Moreover, the structure of LSTM is also helpful for the deep value network to utilize the examined documents. Thus, a deep value network exploits the document feature similarity to promote or demote documents similar or diverse to the previously retrieved and rated documents. 

 To further justify the effect of the feedback method and the ranking model, we design 2 groups of experiments. We replace \myname's ranking model by MDP, which we call it \myname-MDP. Besides, we replace \myname's feedback organizer by the traditional Rocchio algorithm and a navie query expansion method. We denote them as \myname-Rocchio and \myname-nqe. We test them in the Ebola dataset's first 10 queries. The results are shown in Figure 4.

\begin{figure*}[htb]
\centering
\[
\begin{array}{cc}
    \includegraphics[width=220pt]{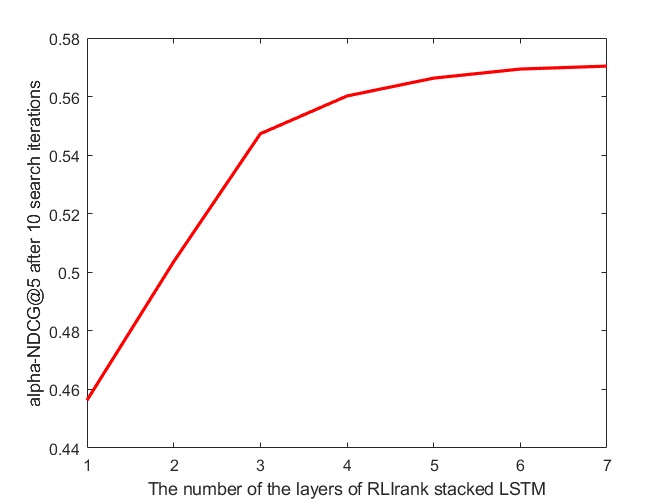}
  &
 \includegraphics[width=220pt]{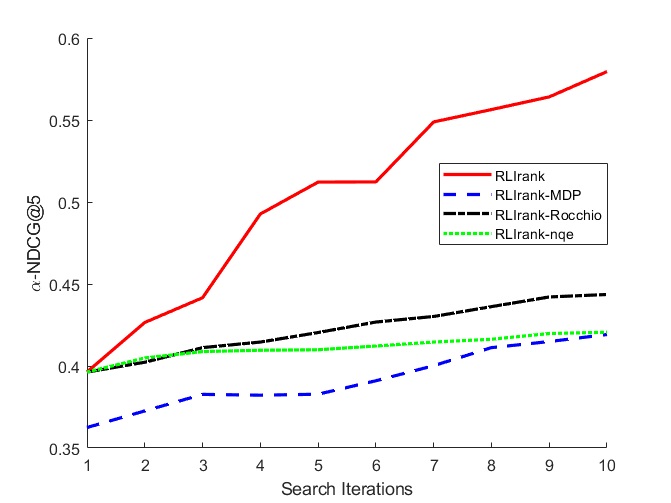}
  \\
  (a) & (b) 
\end{array}
\]
\vspace{-3mm}
\caption{The improvement from the increment of the stacked layers of \myname's LSTM on Ebola 10 (a); $\alpha$-NDCG\@5 of \myname, \myname-MDP, \myname-Rocchio and \myname-nqe on Ebola 10. (b)}
\label{fig:rli-rank-results}
\end{figure*}

Figure 4(b) shows that \myname-MDP has a weaker performance, but improves faster, compared with \myname-Rocchio and \myname-nqe. At the last iteration, the performance of \myname-MDP is competitive. However, \myname outperforms all other variations.  

\section{Conclusions}
We proposed a new dynamic reinforcement-learning based ranking algorithm, \myname{ }, specifically designed for dynamic search where a user provides feedback after each search iteration. 
Additionally, we provide two important contributions for improving the dynamic search: (1) a stacked LSTM trained by a stepwise learning framework, which is the first attempt to incorporate the content (features) of the previously retrieved document into the ranking process. 
(2) An effective embedding Rocchio algorithm, which allows RLIRank to significantly outperform prior state of the art methods for dynamic search.
In summary, our \myname{ } algorithm provides a significant step towards dynamic and contextualized interactive retrieval while opening up promising directions for future work. \\
%

\balance

\bibliographystyle{ACM-Reference-Format}
\bibliography{main}


\begin{thebibliography}{34}


\ifx \showCODEN    \undefined \def \showCODEN     #1{\unskip}     \fi
\ifx \showDOI      \undefined \def \showDOI       #1{#1}\fi
\ifx \showISBNx    \undefined \def \showISBNx     #1{\unskip}     \fi
\ifx \showISBNxiii \undefined \def \showISBNxiii  #1{\unskip}     \fi
\ifx \showISSN     \undefined \def \showISSN      #1{\unskip}     \fi
\ifx \showLCCN     \undefined \def \showLCCN      #1{\unskip}     \fi
\ifx \shownote     \undefined \def \shownote      #1{#1}          \fi
\ifx \showarticletitle \undefined \def \showarticletitle #1{#1}   \fi
\ifx \showURL      \undefined \def \showURL       {\relax}        \fi
\providecommand\bibfield[2]{#2}
\providecommand\bibinfo[2]{#2}
\providecommand\natexlab[1]{#1}
\providecommand\showeprint[2][]{arXiv:#2}

\bibitem[\protect\citeauthoryear{Albahem, Spina, Cavedon, and Scholer}{Albahem
  et~al\mbox{.}}{2016}]%
        {albahem2016rmit}
\bibfield{author}{\bibinfo{person}{Ameer Albahem}, \bibinfo{person}{Damiano
  Spina}, \bibinfo{person}{Lawrence Cavedon}, {and} \bibinfo{person}{Falk
  Scholer}.} \bibinfo{year}{2016}\natexlab{}.
\newblock \showarticletitle{RMIT@ TREC 2016 Dynamic Domain Track: Exploiting
  Passage Representation for Retrieval and Relevance Feedback.}. In
  \bibinfo{booktitle}{\emph{TREC}}.
\newblock


\bibitem[\protect\citeauthoryear{Burges, Shaked, Renshaw, Lazier, Deeds,
  Hamilton, and Hullender}{Burges et~al\mbox{.}}{2005}]%
        {burges2005learning}
\bibfield{author}{\bibinfo{person}{Chris Burges}, \bibinfo{person}{Tal Shaked},
  \bibinfo{person}{Erin Renshaw}, \bibinfo{person}{Ari Lazier},
  \bibinfo{person}{Matt Deeds}, \bibinfo{person}{Nicole Hamilton}, {and}
  \bibinfo{person}{Greg Hullender}.} \bibinfo{year}{2005}\natexlab{}.
\newblock \showarticletitle{Learning to rank using gradient descent}. In
  \bibinfo{booktitle}{\emph{Proceedings of the 22nd international conference on
  Machine learning}}. ACM, \bibinfo{pages}{89--96}.
\newblock


\bibitem[\protect\citeauthoryear{Cao, Qin, Liu, Tsai, and Li}{Cao
  et~al\mbox{.}}{2007}]%
        {cao2007learning}
\bibfield{author}{\bibinfo{person}{Zhe Cao}, \bibinfo{person}{Tao Qin},
  \bibinfo{person}{Tie-Yan Liu}, \bibinfo{person}{Ming-Feng Tsai}, {and}
  \bibinfo{person}{Hang Li}.} \bibinfo{year}{2007}\natexlab{}.
\newblock \showarticletitle{Learning to rank: from pairwise approach to
  listwise approach}. In \bibinfo{booktitle}{\emph{Proceedings of the 24th
  international conference on Machine learning}}. ACM,
  \bibinfo{pages}{129--136}.
\newblock


\bibitem[\protect\citeauthoryear{Cartright, Huston, and Feild}{Cartright
  et~al\mbox{.}}{2012}]%
        {cartright2012galago}
\bibfield{author}{\bibinfo{person}{Marc-Allen Cartright},
  \bibinfo{person}{Samuel Huston}, {and} \bibinfo{person}{Henry Feild}.}
  \bibinfo{year}{2012}\natexlab{}.
\newblock \showarticletitle{Galago: A modular distributed processing and
  retrieval system}. In \bibinfo{booktitle}{\emph{SIGIR 2012 Workshop on Open
  Source Information Retrieval}}. \bibinfo{pages}{25--31}.
\newblock


\bibitem[\protect\citeauthoryear{Clarke, Kolla, Cormack, Vechtomova, Ashkan,
  B{\"u}ttcher, and MacKinnon}{Clarke et~al\mbox{.}}{2008}]%
        {clarke2008novelty}
\bibfield{author}{\bibinfo{person}{Charles~LA Clarke},
  \bibinfo{person}{Maheedhar Kolla}, \bibinfo{person}{Gordon~V Cormack},
  \bibinfo{person}{Olga Vechtomova}, \bibinfo{person}{Azin Ashkan},
  \bibinfo{person}{Stefan B{\"u}ttcher}, {and} \bibinfo{person}{Ian
  MacKinnon}.} \bibinfo{year}{2008}\natexlab{}.
\newblock \showarticletitle{Novelty and diversity in information retrieval
  evaluation}. In \bibinfo{booktitle}{\emph{Proceedings of the 31st annual
  international ACM SIGIR conference on Research and development in information
  retrieval}}. ACM, \bibinfo{pages}{659--666}.
\newblock


\bibitem[\protect\citeauthoryear{Collobert and Weston}{Collobert and
  Weston}{2008}]%
        {collobert2008unified}
\bibfield{author}{\bibinfo{person}{Ronan Collobert} {and}
  \bibinfo{person}{Jason Weston}.} \bibinfo{year}{2008}\natexlab{}.
\newblock \showarticletitle{A unified architecture for natural language
  processing: Deep neural networks with multitask learning}. In
  \bibinfo{booktitle}{\emph{Proceedings of the 25th international conference on
  Machine learning}}. ACM, \bibinfo{pages}{160--167}.
\newblock


\bibitem[\protect\citeauthoryear{Feng, Xu, Lan, Guo, Zeng, and Cheng}{Feng
  et~al\mbox{.}}{2018}]%
        {feng2018greedy}
\bibfield{author}{\bibinfo{person}{Yue Feng}, \bibinfo{person}{Jun Xu},
  \bibinfo{person}{Yanyan Lan}, \bibinfo{person}{Jiafeng Guo},
  \bibinfo{person}{Wei Zeng}, {and} \bibinfo{person}{Xueqi Cheng}.}
  \bibinfo{year}{2018}\natexlab{}.
\newblock \showarticletitle{From Greedy Selection to Exploratory
  Decision-Making: Diverse Ranking with Policy-Value Networks}.
\newblock  (\bibinfo{year}{2018}).
\newblock


\bibitem[\protect\citeauthoryear{Glowacka, Ruotsalo, Konuyshkova, Kaski,
  Jacucci, et~al\mbox{.}}{Glowacka et~al\mbox{.}}{2013}]%
        {glowacka2013directing}
\bibfield{author}{\bibinfo{person}{Dorota Glowacka}, \bibinfo{person}{Tuukka
  Ruotsalo}, \bibinfo{person}{Ksenia Konuyshkova}, \bibinfo{person}{Samuel
  Kaski}, \bibinfo{person}{Giulio Jacucci}, {et~al\mbox{.}}}
  \bibinfo{year}{2013}\natexlab{}.
\newblock \showarticletitle{Directing exploratory search: Reinforcement
  learning from user interactions with keywords}. In
  \bibinfo{booktitle}{\emph{Proceedings of the 2013 international conference on
  Intelligent user interfaces}}. ACM, \bibinfo{pages}{117--128}.
\newblock


\bibitem[\protect\citeauthoryear{Grainger and Potter}{Grainger and
  Potter}{2014}]%
        {grainger2014solr}
\bibfield{author}{\bibinfo{person}{Trey Grainger} {and}
  \bibinfo{person}{Timothy Potter}.} \bibinfo{year}{2014}\natexlab{}.
\newblock \bibinfo{booktitle}{\emph{Solr in action}}.
\newblock \bibinfo{publisher}{Manning Publications Co.}
\newblock


\bibitem[\protect\citeauthoryear{Guan, Zhang, and Yang}{Guan
  et~al\mbox{.}}{2013}]%
        {guan2013utilizing}
\bibfield{author}{\bibinfo{person}{Dongyi Guan}, \bibinfo{person}{Sicong
  Zhang}, {and} \bibinfo{person}{Hui Yang}.} \bibinfo{year}{2013}\natexlab{}.
\newblock \showarticletitle{Utilizing query change for session search}. In
  \bibinfo{booktitle}{\emph{Proceedings of the 36th international ACM SIGIR
  conference on Research and development in information retrieval}}. ACM,
  \bibinfo{pages}{453--462}.
\newblock


\bibitem[\protect\citeauthoryear{Hagen, Stein, and V{\"o}lske}{Hagen
  et~al\mbox{.}}{2010}]%
        {hagen2010webis}
\bibfield{author}{\bibinfo{person}{Matthias Hagen}, \bibinfo{person}{Benno
  Stein}, {and} \bibinfo{person}{Michael V{\"o}lske}.}
  \bibinfo{year}{2010}\natexlab{}.
\newblock \bibinfo{booktitle}{\emph{Webis at the TREC 2010 Sessions track}}.
\newblock \bibinfo{type}{{T}echnical {R}eport}. \bibinfo{institution}{BAUHAUS
  UNIV WEIMAR (GERMANY)}.
\newblock


\bibitem[\protect\citeauthoryear{Joachims}{Joachims}{1996}]%
        {joachims1996probabilistic}
\bibfield{author}{\bibinfo{person}{Thorsten Joachims}.}
  \bibinfo{year}{1996}\natexlab{}.
\newblock \bibinfo{booktitle}{\emph{A Probabilistic Analysis of the Rocchio
  Algorithm with TFIDF for Text Categorization.}}
\newblock \bibinfo{type}{{T}echnical {R}eport}.
  \bibinfo{institution}{Carnegie-mellon univ pittsburgh pa dept of computer
  science}.
\newblock


\bibitem[\protect\citeauthoryear{Kanoulas, Azzopardi, and Yang}{Kanoulas
  et~al\mbox{.}}{2018}]%
        {kanoulas2018overview}
\bibfield{author}{\bibinfo{person}{Evangelos Kanoulas}, \bibinfo{person}{Leif
  Azzopardi}, {and} \bibinfo{person}{Grace~Hui Yang}.}
  \bibinfo{year}{2018}\natexlab{}.
\newblock \showarticletitle{Overview of the CLEF dynamic search evaluation lab
  2018}. In \bibinfo{booktitle}{\emph{International Conference of the
  Cross-Language Evaluation Forum for European Languages}}. Springer,
  \bibinfo{pages}{362--371}.
\newblock


\bibitem[\protect\citeauthoryear{LeCun, Bengio, and Hinton}{LeCun
  et~al\mbox{.}}{2015}]%
        {lecun2015deep}
\bibfield{author}{\bibinfo{person}{Yann LeCun}, \bibinfo{person}{Yoshua
  Bengio}, {and} \bibinfo{person}{Geoffrey Hinton}.}
  \bibinfo{year}{2015}\natexlab{}.
\newblock \showarticletitle{Deep learning}.
\newblock \bibinfo{journal}{\emph{nature}} \bibinfo{volume}{521},
  \bibinfo{number}{7553} (\bibinfo{year}{2015}), \bibinfo{pages}{436}.
\newblock


\bibitem[\protect\citeauthoryear{Lee and Lin}{Lee and Lin}{2014}]%
        {lee2014large}
\bibfield{author}{\bibinfo{person}{Ching-Pei Lee} {and}
  \bibinfo{person}{Chih-Jen Lin}.} \bibinfo{year}{2014}\natexlab{}.
\newblock \showarticletitle{Large-scale linear ranksvm}.
\newblock \bibinfo{journal}{\emph{Neural computation}} \bibinfo{volume}{26},
  \bibinfo{number}{4} (\bibinfo{year}{2014}), \bibinfo{pages}{781--817}.
\newblock


\bibitem[\protect\citeauthoryear{Li, Wu, and Burges}{Li et~al\mbox{.}}{2008}]%
        {li2008mcrank}
\bibfield{author}{\bibinfo{person}{Ping Li}, \bibinfo{person}{Qiang Wu}, {and}
  \bibinfo{person}{Christopher~J Burges}.} \bibinfo{year}{2008}\natexlab{}.
\newblock \showarticletitle{Mcrank: Learning to rank using multiple
  classification and gradient boosting}. In \bibinfo{booktitle}{\emph{Advances
  in neural information processing systems}}. \bibinfo{pages}{897--904}.
\newblock


\bibitem[\protect\citeauthoryear{Li, Dong, Wen, and Guan}{Li
  et~al\mbox{.}}{2018}]%
        {li2018intelligent}
\bibfield{author}{\bibinfo{person}{Yuanlong Li}, \bibinfo{person}{Linsen Dong},
  \bibinfo{person}{Yonggang Wen}, {and} \bibinfo{person}{Kyle Guan}.}
  \bibinfo{year}{2018}\natexlab{}.
\newblock \showarticletitle{Intelligent trainer for model-based reinforcement
  learning}.
\newblock \bibinfo{journal}{\emph{arXiv preprint arXiv:1805.09496}}
  (\bibinfo{year}{2018}).
\newblock


\bibitem[\protect\citeauthoryear{LIN}{LIN}{2007}]%
        {lin2007ordinal}
\bibfield{author}{\bibinfo{person}{HSUAN-TIEN LIN}.}
  \bibinfo{year}{2007}\natexlab{}.
\newblock \showarticletitle{Ordinal regression by extended binary
  classification}.
\newblock \bibinfo{journal}{\emph{Advances in neural information processing
  systems}} (\bibinfo{year}{2007}).
\newblock


\bibitem[\protect\citeauthoryear{Liu, Tang, Li, Ye, Guo, and He}{Liu
  et~al\mbox{.}}{2018}]%
        {liu2018novel}
\bibfield{author}{\bibinfo{person}{Feng Liu}, \bibinfo{person}{Ruiming Tang},
  \bibinfo{person}{Xutao Li}, \bibinfo{person}{Yunming Ye},
  \bibinfo{person}{Huifeng Guo}, {and} \bibinfo{person}{Xiuqiang He}.}
  \bibinfo{year}{2018}\natexlab{}.
\newblock \showarticletitle{Novel Approaches to Accelerating the Convergence
  Rate of Markov Decision Process for Search Result Diversification}. In
  \bibinfo{booktitle}{\emph{International Conference on Database Systems for
  Advanced Applications}}. Springer, \bibinfo{pages}{184--200}.
\newblock


\bibitem[\protect\citeauthoryear{Liu}{Liu}{2009}]%
        {liu2009learning}
\bibfield{author}{\bibinfo{person}{Tie-Yan Liu}.}
  \bibinfo{year}{2009}\natexlab{}.
\newblock \showarticletitle{Learning to rank for information retrieval}.
\newblock \bibinfo{journal}{\emph{Foundations and Trends in Information
  Retrieval}} \bibinfo{volume}{3}, \bibinfo{number}{3} (\bibinfo{year}{2009}),
  \bibinfo{pages}{225--331}.
\newblock


\bibitem[\protect\citeauthoryear{Monahan}{Monahan}{1982}]%
        {monahan1982state}
\bibfield{author}{\bibinfo{person}{George~E Monahan}.}
  \bibinfo{year}{1982}\natexlab{}.
\newblock \showarticletitle{State of the art—a survey of partially observable
  Markov decision processes: theory, models, and algorithms}.
\newblock \bibinfo{journal}{\emph{Management Science}} \bibinfo{volume}{28},
  \bibinfo{number}{1} (\bibinfo{year}{1982}), \bibinfo{pages}{1--16}.
\newblock


\bibitem[\protect\citeauthoryear{Moraes, Santos, and Ziviani}{Moraes
  et~al\mbox{.}}{2016}]%
        {moraes2016ufmg}
\bibfield{author}{\bibinfo{person}{Felipe Moraes}, \bibinfo{person}{Rodrygo~LT
  Santos}, {and} \bibinfo{person}{Nivio Ziviani}.}
  \bibinfo{year}{2016}\natexlab{}.
\newblock \showarticletitle{UFMG at the TREC 2016 Dynamic Domain track.}. In
  \bibinfo{booktitle}{\emph{TREC}}.
\newblock


\bibitem[\protect\citeauthoryear{Sandhaus}{Sandhaus}{2008}]%
        {sandhaus2008new}
\bibfield{author}{\bibinfo{person}{Evan Sandhaus}.}
  \bibinfo{year}{2008}\natexlab{}.
\newblock \showarticletitle{The new york times annotated corpus}.
\newblock \bibinfo{journal}{\emph{Linguistic Data Consortium, Philadelphia}}
  \bibinfo{volume}{6}, \bibinfo{number}{12} (\bibinfo{year}{2008}),
  \bibinfo{pages}{e26752}.
\newblock


\bibitem[\protect\citeauthoryear{Santos, Xiang, and Zhou}{Santos
  et~al\mbox{.}}{2015}]%
        {santos2015classifying}
\bibfield{author}{\bibinfo{person}{Cicero Nogueira~dos Santos},
  \bibinfo{person}{Bing Xiang}, {and} \bibinfo{person}{Bowen Zhou}.}
  \bibinfo{year}{2015}\natexlab{}.
\newblock \showarticletitle{Classifying relations by ranking with convolutional
  neural networks}.
\newblock \bibinfo{journal}{\emph{arXiv preprint arXiv:1504.06580}}
  (\bibinfo{year}{2015}).
\newblock


\bibitem[\protect\citeauthoryear{Schmidhuber}{Schmidhuber}{2015}]%
        {schmidhuber2015deep}
\bibfield{author}{\bibinfo{person}{J{\"u}rgen Schmidhuber}.}
  \bibinfo{year}{2015}\natexlab{}.
\newblock \showarticletitle{Deep learning in neural networks: An overview}.
\newblock \bibinfo{journal}{\emph{Neural networks}}  \bibinfo{volume}{61}
  (\bibinfo{year}{2015}), \bibinfo{pages}{85--117}.
\newblock


\bibitem[\protect\citeauthoryear{Severyn and Moschitti}{Severyn and
  Moschitti}{2015}]%
        {severyn2015learning}
\bibfield{author}{\bibinfo{person}{Aliaksei Severyn} {and}
  \bibinfo{person}{Alessandro Moschitti}.} \bibinfo{year}{2015}\natexlab{}.
\newblock \showarticletitle{Learning to rank short text pairs with
  convolutional deep neural networks}. In \bibinfo{booktitle}{\emph{Proceedings
  of the 38th international ACM SIGIR conference on research and development in
  information retrieval}}. ACM, \bibinfo{pages}{373--382}.
\newblock


\bibitem[\protect\citeauthoryear{Sutskever, Vinyals, and Le}{Sutskever
  et~al\mbox{.}}{2014}]%
        {sutskever2014sequence}
\bibfield{author}{\bibinfo{person}{Ilya Sutskever}, \bibinfo{person}{Oriol
  Vinyals}, {and} \bibinfo{person}{Quoc~V Le}.}
  \bibinfo{year}{2014}\natexlab{}.
\newblock \showarticletitle{Sequence to sequence learning with neural
  networks}. In \bibinfo{booktitle}{\emph{Advances in neural information
  processing systems}}. \bibinfo{pages}{3104--3112}.
\newblock


\bibitem[\protect\citeauthoryear{Sutton and Barto}{Sutton and Barto}{2018}]%
        {sutton2018reinforcement}
\bibfield{author}{\bibinfo{person}{Richard~S Sutton} {and}
  \bibinfo{person}{Andrew~G Barto}.} \bibinfo{year}{2018}\natexlab{}.
\newblock \bibinfo{booktitle}{\emph{Reinforcement learning: An introduction}}.
\newblock \bibinfo{publisher}{MIT press}.
\newblock


\bibitem[\protect\citeauthoryear{Wang, Song, Leung, Rosenberg, Wang, Philbin,
  Chen, and Wu}{Wang et~al\mbox{.}}{2014}]%
        {wang2014learning}
\bibfield{author}{\bibinfo{person}{Jiang Wang}, \bibinfo{person}{Yang Song},
  \bibinfo{person}{Thomas Leung}, \bibinfo{person}{Chuck Rosenberg},
  \bibinfo{person}{Jingbin Wang}, \bibinfo{person}{James Philbin},
  \bibinfo{person}{Bo Chen}, {and} \bibinfo{person}{Ying Wu}.}
  \bibinfo{year}{2014}\natexlab{}.
\newblock \showarticletitle{Learning fine-grained image similarity with deep
  ranking}. In \bibinfo{booktitle}{\emph{Proceedings of the IEEE Conference on
  Computer Vision and Pattern Recognition}}. \bibinfo{pages}{1386--1393}.
\newblock


\bibitem[\protect\citeauthoryear{Wei, Xu, Lan, Guo, and Cheng}{Wei
  et~al\mbox{.}}{2017}]%
        {wei2017reinforcement}
\bibfield{author}{\bibinfo{person}{Zeng Wei}, \bibinfo{person}{Jun Xu},
  \bibinfo{person}{Yanyan Lan}, \bibinfo{person}{Jiafeng Guo}, {and}
  \bibinfo{person}{Xueqi Cheng}.} \bibinfo{year}{2017}\natexlab{}.
\newblock \showarticletitle{Reinforcement learning to rank with Markov decision
  process}. In \bibinfo{booktitle}{\emph{Proceedings of the 40th International
  ACM SIGIR Conference on Research and Development in Information Retrieval}}.
  ACM, \bibinfo{pages}{945--948}.
\newblock


\bibitem[\protect\citeauthoryear{Xu and Li}{Xu and Li}{2007}]%
        {xu2007adarank}
\bibfield{author}{\bibinfo{person}{Jun Xu} {and} \bibinfo{person}{Hang Li}.}
  \bibinfo{year}{2007}\natexlab{}.
\newblock \showarticletitle{Adarank: a boosting algorithm for information
  retrieval}. In \bibinfo{booktitle}{\emph{Proceedings of the 30th annual
  international ACM SIGIR conference on Research and development in information
  retrieval}}. ACM, \bibinfo{pages}{391--398}.
\newblock


\bibitem[\protect\citeauthoryear{Yang and Yang}{Yang and Yang}{2017}]%
        {yang2017contextual}
\bibfield{author}{\bibinfo{person}{Angela Yang} {and}
  \bibinfo{person}{Grace~Hui Yang}.} \bibinfo{year}{2017}\natexlab{}.
\newblock \showarticletitle{A contextual bandit approach to dynamic search}. In
  \bibinfo{booktitle}{\emph{Proceedings of the ACM SIGIR International
  Conference on Theory of Information Retrieval}}. ACM,
  \bibinfo{pages}{301--304}.
\newblock


\bibitem[\protect\citeauthoryear{Yang and Soboroff}{Yang and Soboroff}{2016}]%
        {yang2016trec}
\bibfield{author}{\bibinfo{person}{Grace~Hui Yang} {and} \bibinfo{person}{Ian
  Soboroff}.} \bibinfo{year}{2016}\natexlab{}.
\newblock \showarticletitle{TREC 2016 Dynamic Domain Track Overview.}. In
  \bibinfo{booktitle}{\emph{TREC}}.
\newblock


\bibitem[\protect\citeauthoryear{Zhang, Hu, Jia, Wang, Zhang, Feng, Yu, Xue,
  Yu, Liu, et~al\mbox{.}}{Zhang et~al\mbox{.}}{2017}]%
        {zhang2017ictnet}
\bibfield{author}{\bibinfo{person}{Weimin Zhang}, \bibinfo{person}{Yaokang Hu},
  \bibinfo{person}{Rongqian Jia}, \bibinfo{person}{Xianfa Wang},
  \bibinfo{person}{Le Zhang}, \bibinfo{person}{Yue Feng},
  \bibinfo{person}{Sihao Yu}, \bibinfo{person}{Yuanhai Xue},
  \bibinfo{person}{Xiaoming Yu}, \bibinfo{person}{Yue Liu}, {et~al\mbox{.}}}
  \bibinfo{year}{2017}\natexlab{}.
\newblock \showarticletitle{ICTNET at TREC 2017 Dynamic Domain Track.}. In
  \bibinfo{booktitle}{\emph{TREC}}.
\newblock


\end{thebibliography}

%
\appendix

\end{document}